\documentclass[a4paper, 11pt]{article}
\pdfoutput=1

\usepackage{jheppub}

\newcommand{\eVq}{\ensuremath{\text{eV}^2}}
\newcommand{\Dmq}{\Delta m^2}
\newcommand{\Eps}{\varepsilon}
\newcommand{\dltCP}{\delta_\text{CP}}
\newcommand{\diag}{\mathop{\mathrm{diag}}}
\renewcommand{\Re}{\mathop{\mathrm{Re}}}
\newcommand{\Nuc}[2]{\ensuremath{\mbox{}^{#1}\text{#2}}}

\allowdisplaybreaks

\title{Determination of matter potential from global analysis of
  neutrino oscillation data}

\author[a,b]{M.~C.~Gonzalez-Garcia,}
\affiliation[a]{C.N.~Yang Institute for Theoretical Physics, State
  University of New York at Stony Brook, Stony Brook, NY 11794-3840,
  USA}
\affiliation[b]{Instituci\'o Catalana de Recerca i Estudis
  Avan\c{c}ats (ICREA), Departament d'Estructura i Constituents de la
  Mat\`eria and Institut de Ciencies del Cosmos, Universitat de
  Barcelona, Diagonal 647, E-08028 Barcelona, Spain}
\emailAdd{concha@insti.physics.sunysb.edu}

\author[c]{Michele Maltoni,}
\affiliation[c]{Instituto de F\'{\i}sica Te\'orica UAM/CSIC, Calle de
  Nicol\'as Cabrera 13--15, Universidad Aut\'onoma de Madrid,
  Cantoblanco, E-28049 Madrid, Spain}
\emailAdd{michele.maltoni@csic.es}

\abstract{We quantify our current knowledge of the size and flavor
  structure of the matter effects in the evolution of neutrinos based
  solely on the global analysis of oscillation neutrino data. The
  results are translated in terms of the present allowed ranges for
  the corresponding non-standard neutrino interactions in matter.}


\preprint{IFT-UAM/CSIC-13-078, YITP-SB-13-20}

\begin{document}

\maketitle

\section{Introduction}

It is now an established fact that neutrinos are massive and leptonic
flavors are not symmetries of Nature~\cite{Pontecorvo:1967fh,
  Gribov:1968kq}.  This picture has now become fully proved thanks to
the upcoming of a set of precise experiments which have confirmed the
results obtained with solar and atmospheric neutrinos using
terrestrial beams of neutrinos produced in nuclear reactors and
accelerators facilities~\cite{GonzalezGarcia:2007ib}.
The minimum joint description of the neutrino data requires mixing
among all the three known neutrinos ($\nu_e$, $\nu_\mu$, $\nu_\tau$),
which can be expressed as quantum superposition of three massive
states $\nu_i$ ($i=1,2,3$) with masses $m_i$. Consequently when
written in terms of mass eigenstates, the weak charged current
interactions of leptons~\cite{Maki:1962mu, Kobayashi:1973fv} contain a
leptonic mixing matrix which can be parametrized as:
\begin{equation}
  \label{eq:matrix}
  U_\text{vac} =
  \begin{pmatrix}
    c_{12} c_{13}
    & s_{12} c_{13}
    & s_{13} e^{-i\dltCP}
    \\
    - s_{12} c_{23} - c_{12} s_{13} s_{23} e^{i\dltCP}
    & \hphantom{+} c_{12} c_{23} - s_{12} s_{13} s_{23} e^{i\dltCP}
    & c_{13} s_{23} \hspace*{5.5mm}
    \\
    \hphantom{+} s_{12} s_{23} - c_{12} s_{13} c_{23} e^{i\dltCP}
    & - c_{12} s_{23} - s_{12} s_{13} c_{23} e^{i\dltCP}
    & c_{13} c_{23} \hspace*{5.5mm}
  \end{pmatrix},
\end{equation}
where $c_{ij} \equiv \cos\theta_{ij}$ and $s_{ij} \equiv
\sin\theta_{ij}$.  In addition to the Dirac-type phase $\dltCP$,
analogous to that of the quark sector, there are two physical phases
associated to the Majorana character of neutrinos, which are not
relevant for neutrino oscillations~\cite{Bilenky:1980cx,
  Langacker:1986jv} and which are therefore omitted in the following.

In the simplest quantum-mechanical picture, flavor oscillations are
generated by the kinematical Hamiltonian for this ensemble, 
$H_\text{vac}$, which in the flavor basis $(\nu_e, \nu_\mu, \nu_\tau)$
reads
\begin{equation}
  \label{eq:hvac}
  H_\text{vac} = U_\text{vac} D_\text{vac} U_\text{vac}^\dagger
  \quad\text{with}\quad
  D_\text{vac} = \frac{1}{2E_\nu} \diag(0, \Dmq_{21}, \Dmq_{31})
\end{equation}
The quantities $\Dmq_{21}$, $|\Dmq_{31}|$, $\theta_{12}$,
$\theta_{23}$, and $\theta_{13}$ are relatively well determined by the
analysis of solar, atmospheric, reactor and accelerator experiments,
while barely nothing is known on the CP phase $\dltCP$ and on the sign
of $\Dmq_{31}$~\cite{nufit-1.1, GonzalezGarcia:2012sz, Fogli:2012ua,
  Tortola:2012te}.
Given the observed hierarchy between the solar and atmospheric
mass-squared splittings there are two possible non-equivalent
orderings for the mass eigenvalues, which are conventionally chosen as
\begin{align}
  \label{eq:normal}
  \Dmq_{21} &\ll (\Dmq_{32} \simeq \Dmq_{31})
  \text{ with } (\Dmq_{31} > 0) \,;
  \\
  \label{eq:inverted}
  \Dmq_{21} &\ll |\Dmq_{31} \simeq \Dmq_{32}|
  \text{ with } (\Dmq_{31} < 0) \,.
\end{align}
As it is customary we refer to the first option,
Eq.~\eqref{eq:normal}, as the \emph{normal} ordering, and to the
second one, Eq.~\eqref{eq:inverted}, as the \emph{inverted}
ordering. Clearly they correspond to the two possible choices of the
sign of $\Dmq_{31}$.

The flavor evolution of this neutrino ensemble is also affected by the
difference in the matter potential induced by neutrino-matter
interactions through the so-called Mikheev-Smirnov-Wolfenstein (MSW)
mechanism~\cite{Wolfenstein:1977ue, Mikheev:1986gs}.  Within the
context of the Standard Model (SM) of particle interactions, this
effect is fully determined and leads to a matter potential which, for
neutral matter, is proportional to the number density of electrons in
the background $N_e(r)$, $V=\sqrt{2} G_F N_e(r)$, and which only
affects electron neutrinos. The evolution of the ensemble is then
determined by the Hamiltonian $ H^\nu = H_\text{vac} +
H^\text{SM}_\text{mat}$, with $H^\text{SM}_\text{mat} = \sqrt{2} G_F
N_e(r) \diag(1, 0, 0)$.  The magnitude and the presence of
non-standard forms of the matter potential can be tested in solar
neutrino experiments (and in combination with
KamLAND)~\cite{Roulet:1991sm, Guzzo:1991hi, Barger:1991ae,
  Fogli:1993xv, Bergmann:1997mr, Bergmann:2000gp, Guzzo:2000kx,
  Fogli:2002hb, Friedland:2004pp, Escrihuela:2009up, Bolanos:2008km,
  Minakata:2010be, Palazzo:2011vg, Bonventre:2013loa}, as well and in
the propagation of atmospheric and long-baseline
neutrinos~\cite{Grossman:1995wx, GonzalezGarcia:2001mp, Gago:2001xg,
  Fornengo:2001pm, Huber:2001zw, Ota:2001pw, Huber:2002bi,
  Campanelli:2002cc, Ota:2002na, GonzalezGarcia:2004wg,
  Friedland:2004ah, Friedland:2005vy, Blennow:2005qj, Kitazawa:2006iq,
  Friedland:2006pi, Blennow:2007pu, Kopp:2007mi, Kopp:2007ne,
  Ribeiro:2007ud, Bandyopadhyay:2007kx, Ribeiro:2007jq,
  EstebanPretel:2008qi, Blennow:2008ym, Kopp:2008ds, Ohlsson:2008gx,
  Palazzo:2009rb, GonzalezGarcia:2011my}.

In this article we address our current knowledge of the size and
flavor structure of the matter background effects in the evolution of
solar, atmospheric, reactor and long-baseline (LBL) accelerator
neutrinos based on the global analysis of oscillation data.  To this
aim, in Sec.~\ref{sec:formalism} we briefly present the most general
parametrization of the matter potential and its connection with
non-standard neutrino interactions (NSI) in matter, which provide a
well-known theoretical framework for this kind of phenomenological
studies. We also discuss the simplifications used in the analysis of
the solar+KamLAND sector and the atmospheric+LBL sector
respectively. In Sec.~\ref{sec:solar} we present the results from the
updated analysis of solar+KamLAND data and quantify the impact of the
modified matter potential on the data description, as well as the
status of the well-known ``dark-side'' solution which appears in
presence of NSI.  In Ref.~\cite{GonzalezGarcia:2011my} an analysis of
atmospheric and LBL neutrino data was performed in the framework of a
generalized matter potential, which extended the standard one by
allowing for an arbitrary rescaling of the potential strength, a
general rotation from the $ee$ sector, and a rephasing with respect to
$H_\text{vac}$. It was concluded that the strength of the potential
cannot be determined solely by these data, whereas its flavor
composition is very much constrained. In Sec.~\ref{sec:global} we
update this analysis and revisit its conclusions after combining the
results from atmospheric, LBL and reactor experiments with those from
solar+KamLAND data.  We show to what degree the determination of
neutrino masses and mixing is robust even in the presence of this
general form of the matter potential and we derive the most up-to-date
allowed ranges on NSI parameters. Finally in Sec.~\ref{sec:summary} we
summarize our results.

\section{Formalism}
\label{sec:formalism}

In the three-flavor oscillation picture, the neutrino evolution
equation reads:
\begin{equation}
  i\frac{d}{dx}
  \begin{pmatrix}
    \nu_e\\
    \nu_\mu\\
    \nu_\tau
  \end{pmatrix}
  = H^\nu
  \begin{pmatrix}
    \nu_e\\
    \nu_\mu\\
    \nu_\tau
  \end{pmatrix}
\end{equation}
where $x$ is the coordinate along the neutrino trajectory and the
Hamiltonian for neutrinos and antineutrinos is:
\begin{equation}
  H^\nu = H_\text{vac} + H_\text{mat}
  \quad\text{and}\quad
  H^{\bar\nu} = ( H_\text{vac} - H_\text{mat} )^* \,,
\end{equation}
with $ H_\text{vac}$ given in Eq.~\eqref{eq:hvac}. Thus the vacuum
term has $6$ parameters: $\Dmq_{21}$, $\Dmq_{31}$, $\theta_{12}$,
$\theta_{13}$, $\theta_{23}$, $\dltCP$.
In the Standard Model $H_\text{mat}$ is fully determined both in its
strength and flavor structure to be $H^\text{SM}_\text{mat} =\sqrt{2}
G_F N_e(r) \diag(1, 0, 0)$ for ordinary matter.  Generically ordinary
matter is composed by electrons ($e$), up-quarks ($u$) and down-quarks
($d$), thus in the most general case a non-standard matter potential
can be parametrized as:
\begin{equation}
  \label{eq:hmatNSI}
  H_\text{mat} = \sqrt{2} G_F N_e(r)
  \begin{pmatrix}
    1 & 0 & 0 \\
    0 & 0 & 0 \\
    0 & 0 & 0
  \end{pmatrix}
  + \sqrt{2} G_F \sum_{f=e,u,d} N_f(r)
  \begin{pmatrix}
    \Eps_{ee}^f & \Eps_{e\mu}^f & \Eps_{e\tau}^f
    \\
    \Eps_{e\mu}^{f*} & \Eps_{\mu\mu}^f & \Eps_{\mu\tau}^f
    \\
    \Eps_{e\tau}^{f*} & \Eps_{\mu\tau}^{f*} & \Eps_{\tau\tau}^f
  \end{pmatrix} .
\end{equation}
Since this matter term can be determined by oscillation experiments
only up to an overall multiple of the identity, without loss of
generality one can assume $\Eps_{\mu\mu}^f = 0$. With this, we have 8
parameters (for each $f$) since $\Eps_{ee}^f$ and $\Eps_{\tau\tau}^f$
must be real whereas $\Eps_{e\mu}^f$, $\Eps_{e\tau}^f$ and
$\Eps_{\mu\tau}^f$ can be complex.

In order to determine the relevant ranges for the parameters in the
problem we must study which transformations leave the probabilities
invariant. In particular we notice that any rephasing $H^\nu \to Q H
Q^*$ where $Q = \diag\left( e^{ia}, e^{ib}, e^{ic} \right)$ leads to a
rephasing of the scattering matrix $\exp(-iH^\nu L) \to Q \exp(-iH^\nu
L) Q^*$, which does not affect the probabilities.  In the standard
oscillation scenario these symmetries are used to reduce the range of
the mixing parameters, most commonly to $0 \leq \theta_{ij} \leq
\pi/2$ and $0 \leq \dltCP \leq 2\pi$.  In the presence of the
non-standard matter potential they can be used just in the same way,
thus reducing the range of the mixing parameters while keeping the
phases of all the off-diagonal $\Eps_{\alpha\neq\beta}^f$.
Alternatively, one could instead reduce the range for some of the
$\Eps_{\alpha\neq\beta}^f$, at the price of retaining a wider range of
the vacuum mixing angles.  Furthermore, in the particular case of a
unique $f$ and in the absence of the vacuum term it would be possible
to use these symmetries to reduce the matter potential parameters from
eight to six: two real flavor diagonal parameters, the absolute value
of the flavor off-diagonal parameters, $|\Eps_{\alpha\neq\beta}^f|$,
and one combination of their three complex phases, while the two
additional phases would become unphysical.  Only when both the vacuum
term and the matter general potential are present the two additional
phases become observable. Hence it is clear from this discussion that
it is a matter of convention to include them in the matter potential
or in the vacuum term.  For real matter potential this means that only
an overall sign of the three off-diagonal $\Eps_{\alpha\neq\beta}^f$
can be considered a generic feature of the matter potential, while the
other two signs of $\Eps_{\alpha\neq\beta}^f$ can be traded off by
enlarging the vacuum mixing parameters to $-\pi/2 \leq \theta_{ij}
\leq \pi/2$.  We will go back to this issue in the next section.
 
The standard theoretical framework for our proposed parametrization of
the matter potential is provided by NSI affecting neutrino
interactions in matter.  They can be described by effective
four-fermion operators of the form
\begin{equation}
  \label{eq:def}
  \mathcal{L}_\text{NSI} =
  - 2\sqrt{2} G_F \Eps_{\alpha\beta}^{fP}
  (\bar\nu_{\alpha} \gamma^\mu \nu_{\beta})
  (\bar{f} \gamma_\mu P f) \,,
\end{equation}
where $f$ is a charged fermion, $P=(L,R)$ and
$\Eps_{\alpha\beta}^{fP}$ are dimensionless parameters encoding the
deviation from standard interactions.  NSI enter in neutrino
propagation only through the vector couplings so the induced matter
Hamiltonian takes the form~\eqref{eq:hmatNSI} with
$\Eps_{\alpha\beta}^f = \Eps_{\alpha\beta}^{fL} +
\Eps_{\alpha\beta}^{fR}$.

\subsection{Earth matter potential for atmospheric and LBL neutrinos}
\label{sec:form-atmos}

As seen above, in principle a generalized potential involves different
parameters for the different charged fermions $f=e,u,d$ in the matter.
In practice, however, for the propagation of atmospheric and LBL
neutrinos the neutron/electron ratio $Y_n$ is reasonably constant all
over the Earth. This implies that neutrino atmospheric and LBL
oscillations are only sensitive to the \emph{sum} of these
interactions, weighted with the relative abundance of each
particle. We can therefore define:
\begin{equation}
  \label{eq:compoconst}
  \Eps_{\alpha\beta} \equiv
  \sum_{f=e,u,d} \left< \frac{Y_f}{Y_e} \right> \Eps_{\alpha\beta}^f
  = \Eps_{\alpha\beta}^e + Y_u\, \Eps_{\alpha\beta}^u + Y_d\, \Eps_{\alpha\beta}^d
\end{equation}
The PREM model~\cite{Dziewonski:1981xy} fixes $Y_n = 1.012$ in the
Mantle and $Y_n = 1.137$ in the Core, with an average value $Y_n =
1.051$ all over the Earth. Since a proton has 2 up-quarks and 1
down-quark, a neutron has 1 up-quark and 2 down-quarks, and neutral
matter obviously has the same number of protons and electrons ($Y_p =
1$), we get $Y_u = 2 + Y_n = 3.051$ and $Y_d = 1 + 2 Y_n = 3.102$ in
the Earth. With this in mind, the matter part of the Hamiltonian can
be written as:
\begin{equation}
  \label{eq:hmatNSIatm}
  H_\text{mat} = \sqrt{2} G_F N_e(r)
  \begin{pmatrix}
    1 + \Eps_{ee} & \Eps_{e\mu} & \Eps_{e\tau}
    \\
    \Eps_{e\mu}^* & \Eps_{\mu\mu} & \Eps_{\mu\tau}
    \\
    \Eps_{e\tau}^* & \Eps_{\mu\tau}^* & \Eps_{\tau\tau}
  \end{pmatrix}
\end{equation}
where the standard interactions are accounted by the ``$1\,+$'' term
in the $ee$ entry, and the non-standard interactions are accounted by
the $\Eps_{\alpha\beta}$ terms.
Since $H_\text{mat}$ is Hermitian and its trace is irrelevant for
oscillations, we have 8 parameters.

In Ref.~\cite{GonzalezGarcia:2011my} an alternative parametrization
for $H_\text{mat}$, mimicking the structure of the vacuum term in
Eq.~\eqref{eq:hvac}, was introduced as
\begin{equation}
  \label{eq:hmatgen}
  H_\text{mat} = Q_\text{rel} U_\text{mat} D_\text{mat}
  U_\text{mat}^\dagger Q_\text{rel}^\dagger
  \text{~~with~~}
  \left\lbrace
  \begin{aligned}
    Q_\text{rel} &= \diag\left(
    e^{i\alpha_1}, e^{i\alpha_2}, e^{-i\alpha_1 -i\alpha_2} \right) ,
    \\
    U_\text{mat} &= R_{12}(\varphi_{12})
    \tilde{R}_{13}(\varphi_{13}, \delta_\text{NS})
    R_{23}(\varphi_{23}) \,,
    \\
    D_\text{mat} &= \sqrt{2} G_F N_e(r) \diag(\Eps, \Eps', 0)
  \end{aligned}\right.
\end{equation}
where we denote by $R_{ij}(\varphi_{ij})$ a rotation of angle
$\varphi_{ij}$ in the $ij$ plane and
$\tilde{R}_{13}(\varphi_{13},\delta_\text{NS})$ is a \emph{complex}
rotation by angle $\psi_{13}$ and phase $\delta_\text{NS}$.  Just as
Eq.~\eqref{eq:hmatgen} this parametrization also contains 8 real
parameters: 2 eigenvalues, 3 angles and 3 phases. The two phases
$\alpha_1$ and $\alpha_2$ included in $Q_\text{rel}$ are not a feature
of neutrino-matter interactions, but rather a relative feature of the
vacuum and matter term: they would become unphysical if any of the two
terms were not there. Reinterpreted in the notation of
Eq.~\eqref{eq:hmatNSIatm}, this means that only one particular
combination of the three complex phases of $\Eps_{e\mu}$,
$\Eps_{e\tau}$, $\Eps_{\mu\tau}$ is a genuine property of NSI.  In
other words, the relation in Eq.~\eqref{eq:compoconst} implies that
the matter potential behaves as composed of a unique effective
fermion, and in this case, as discussed in the previous section, it is
a matter of convention to define the off-diagonal elements of the
matter potential as three complex parameters, or as three positive
real parameters plus a matter CP phase, and the two additional phases
being assigned to either vacuum or matter part.

Further simplification follows from neglecting $\Dmq_{21}$ in the
analysis of atmospheric, LBL and all reactor experiments but KamLAND,
and by imposing that two eigenvalues of the $H_\text{mat}$ are equal
($\Eps'=0$).  In the limit $\Dmq_{21} \to 0$ the $\theta_{12}$ angle
and the $\dltCP$ phase become unphysical, even in the presence of the
generalized $H_\text{mat}$ in Eq.~\eqref{eq:hmatgen}.
Similarly, for $\Eps' \to 0$ the $\varphi_{23}$ angle and the
$\delta_\text{NS}$ phase become unphysical and the general
$H_\text{mat}$ contains 5 real parameters: $\Eps$ which represents a
rescaling of the matter potential strength, $\varphi_{12}$ and
$\varphi_{13}$ which allows for projection of the potential into the
$\nu_\mu$ and $\nu_\tau$ flavors, and the 2 vacuum-matter relative
phases $\alpha_1$ and $\alpha_2$.
In Ref.~\cite{Friedland:2004ah} it was shown that strong cancellations
in the oscillation of atmospheric neutrinos occur when two eigenvalues
of $H_\text{mat}$ are equal, so that although the limit $\Eps' = 0$
considered here is only a subspace of the most general case on
non-standard interactions, it is precisely in this subspace where the
weakest constraints can be placed.
Under these assumptions the relations between the original
$\Eps_{\alpha\beta}$ in Eq.~\eqref{eq:hmatNSIatm} and the parameters
in Eq.~\eqref{eq:hmatgen} read:
\begin{equation}
  \label{eq:eps_atm}
  \begin{aligned}
    \Eps_{ee} - \Eps_{\mu\mu}
    &= \hphantom{-} \Eps \, (\cos^2\varphi_{12} - \sin^2\varphi_{12})
    \cos^2\varphi_{13} - 1\,,
    \\
    \Eps_{\tau\tau} - \Eps_{\mu\mu}
    &= \hphantom{-} \Eps \, (\sin^2\varphi_{13}
    - \sin^2\varphi_{12} \, \cos^2\varphi_{13}) \,,
    \\
    \Eps_{e\mu}
    &= -\Eps \, \cos\varphi_{12} \, \sin\varphi_{12} \,
    \cos^2\varphi_{13} \, e^{i(\alpha_1 - \alpha_2)} \,,
    \\
    \Eps_{e\tau}
    &= -\Eps \, \cos\varphi_{12} \, \cos\varphi_{13} \,
    \sin\varphi_{13} \, e^{i(2\alpha_1 + \alpha_2)} \,,
    \\
    \Eps_{\mu\tau}
    &= \hphantom{-} \Eps \, \sin\varphi_{12} \, \cos\varphi_{13} \,
    \sin\varphi_{13} \, e^{i(\alpha_1 + 2\alpha_2)} \,,
  \end{aligned}
\end{equation}
which makes explicit that the diagonal terms ($\Eps_{ee}$,
$\Eps_{\mu\mu}$, $\Eps_{\tau\tau}$) can only be determined up to an
overall additive constant. The term ``$-\,1$'' at the end of
$\Eps_{ee} - \Eps_{\mu\mu}$ arises from the standard matter term. The
fermion-specific coefficients $\Eps_{\alpha\beta}^f$ are obtained from
the effective ones $\Eps_{\alpha\beta}$ just by rescaling:
\begin{equation}
  \label{eq:eps_true}
  \Eps_{\alpha\beta}^e = \Eps_{\alpha\beta} \,,
  \qquad
  \Eps_{\alpha\beta}^u = \Eps_{\alpha\beta} \big/ Y_u \,,
  \qquad
  \Eps_{\alpha\beta}^d = \Eps_{\alpha\beta} \big/ Y_d \,.
\end{equation}
Thus, in summary, the relevant flavor transition probabilities for
atmospheric and LBL experiments depend on eight parameters:
($\Dmq_{31}$, $\theta_{13}$, $\theta_{23}$) for the vacuum part,
($\Eps$, $\varphi_{12}$, $\varphi_{13}$) for the matter part, and
($\alpha_1$, $\alpha_2$) as relative phases. As for reactor
experiments other than KamLAND, matter effects are completely
irrelevant due to the very small amount of matter crossed, so the
corresponding $P_{ee}$ survival probability only depends on the two
parameters ($\Dmq_{31}$, $\theta_{13}$).

As shown in Appendix B of Ref.~\cite{GonzalezGarcia:2011my}, only the
relative sign of $\Dmq_{31}$ and $\Eps$ is relevant for atmospheric
and LBL neutrino oscillations. Concerning the angles, in the general
case of unconstrained $\alpha_i$ it is enough to consider $0 <
\theta_{ij} < \pi/2$ and $0 < \varphi_{ij} < \pi/2$, whereas for the
case of \emph{real} NSI (corresponding to $\alpha_i \in \lbrace 0,\pi
\rbrace$) we can set $\alpha_1 = \alpha_2 = 0$ and extend the
$\varphi_{ij}$ range to $-\pi/2 < \varphi_{ij} < \pi/2$.

\subsection{Earth matter potential for solar and KamLAND neutrinos}
\label{sec:form-solar}

For the study of propagation of solar and KamLAND neutrinos one can
work in the one mass dominance approximation, $\Dmq_{31} \to \infty$
(which effectively means that generically $G_F \sum_f N_f(r)
\Eps_{\alpha\beta}^f \ll \Dmq_{31} / E_\nu$). In this approximation
the survival probability $P_{ee}$ can be written as~\cite{Kuo:1986sk,
  Guzzo:2000kx}
\begin{equation}
  \label{eq:peesun}
  P_{ee} = c_{13}^4 P_\text{eff} + s_{13}^4
\end{equation}
where $c_{ij} \equiv \cos\theta_{ij}$ and $s_{ij} \equiv
\sin\theta_{ij}$. The probability $P_\text{eff}$ can be calculated in
an effective $2\times 2$ model with the Hamiltonian $H_\text{eff} =
H_\text{vac}^\text{eff} + H_\text{mat}^\text{eff}$, where:
\begin{align}
  \label{eq:hvacsol}
  H_\text{vac}^\text{eff}
  &= \frac{\Dmq_{21}}{4 E_\nu}
  \begin{pmatrix}
    -\cos 2\theta_{12} & \sin 2\theta_{12} \\
    \hphantom{+} \sin 2\theta_{12} & \cos 2\theta_{12}
  \end{pmatrix} ,
  \\
  \label{eq:hmatsol}
  H_\text{mat}^\text{eff}
  &= \sqrt{2} G_F N_e(r)
  \begin{pmatrix}
    c_{13}^2 & 0 \\
    0 & 0
  \end{pmatrix}
  + \sqrt{2} G_F \sum_f N_f(r)
  \begin{pmatrix}
    -\Eps_D^{f\hphantom{*}} & \Eps_N^f \\
    \hphantom{+} \Eps_N^{f*} & \Eps_D^f
  \end{pmatrix} .
\end{align}
The coefficients $\Eps_D^f$ and $\Eps_N^f$ are related to the original
parameters $\Eps_{\alpha\beta}^f$ by the following relations:
\begin{align}
  \label{eq:eps_D}
  \begin{split}
    \Eps_D^f &=
    c_{13} s_{13} \Re\left[ e^{i\dltCP} \big( s_{23} \, \Eps_{e\mu}^f
      + c_{23} \, \Eps_{e\tau}^f \big) \right]
    - \big( 1 + s_{13}^2 \big) c_{23} s_{23} \Re\!\big( \Eps_{\mu\tau}^f \big)
    \\
    & \hphantom{={}}
    -\frac{c_{13}^2}{2} \big( \Eps_{ee}^f - \Eps_{\mu\mu}^f \big)
    + \frac{s_{23}^2 - s_{13}^2 c_{23}^2}{2}
    \big( \Eps_{\tau\tau}^f - \Eps_{\mu\mu}^f \big) \,,
  \end{split}
  \\[2mm]
  \label{eq:eps_N}
  \Eps_N^f &=
  c_{13} \big( c_{23} \, \Eps_{e\mu}^f - s_{23} \, \Eps_{e\tau}^f \big)
  + s_{13} e^{-i\dltCP} \left[
    s_{23}^2 \, \Eps_{\mu\tau}^f - c_{23}^2 \, \Eps_{\mu\tau}^{f*}
    + c_{23} s_{23} \big( \Eps_{\tau\tau}^f - \Eps_{\mu\mu}^f \big)
    \right] ,
\end{align}
so effectively the relevant probabilities for solar and KamLAND
neutrinos depend on the 3 real oscillation parameters $\Dmq_{21}$,
$\theta_{12}$, and $\theta_{13}$ as well as one real $\Eps_D^f$ and
one complex $\Eps_N^f$ matter parameter for each $f$.  Notice also
that the matter chemical composition of the Sun varies substantially
along the neutrino production region, with $Y_n$ dropping from about
$1/2$ in the center to about $1/6$ at the border of the solar
core. Therefore, unlike the case of Eq.~\eqref{eq:compoconst} for the
Earth it is not possible to introduce a common set of parameters
accounting simultaneously for all the different $f$.  Consequently in
the analysis of solar data we will consider only one particular choice
of $f=e$, $f=u$ or $f=d$ at a time.

Concerning the parameter ranges, the situation is very similar to the
standard case without NSI. The angle $\theta_{13}$ only enters through
Eq.~\eqref{eq:peesun}, so it is sufficient to consider $0 \le
\theta_{13} \le \pi/2$. The Hamiltonian~\eqref{eq:hmatsol} is
invariant under the transformation $\Dmq_{21} \to -\Dmq_{21}$ $\wedge$
$\theta_{12} \to \theta_{12} + \pi/2$, so without loss of generality
we can assume $\Dmq_{21} > 0$. In Eq.~\eqref{eq:hvacsol} $\theta_{12}$
appears multiplied by $2$, so we can restrict its range to $-\pi/2 \le
\theta_{12} \le +\pi/2$. Finally, the probabilities are insensitive to
the overall sign of the non-diagonal entry of \eqref{eq:hmatsol},
resulting in a symmetry $\theta_{12} \to -\theta_{12}$ $\wedge$
$\Eps_N^f \to -\Eps_N^f$, which can be used to further restrict the
$\theta_{12}$ range to $0 \le \theta_{12} \le \pi/2$. Thus in the most
general case we have $\Dmq_{21} > 0$, $0 \le \theta_{ij} \le \pi/2$,
$\Eps_D^f$ real, and $\Eps_N^f$ complex. Notice however that, as
discussed before, from the point of view of neutrino oscillations the
phase of $\Eps_N^f$ is not a genuine NSI property but rather a
relative feature of the vacuum and matter parts.

In the specific case of non-standard interactions with
\emph{electrons} ($f=e$) there is another exact symmetry. Both the
standard and the non-standard terms in Eq.~\eqref{eq:hmatsol} scale
with the same matter density profile $N_e(r)$, so they can be merged
into a single term and $H_\text{mat}^\text{eff}$ takes the form:
\begin{equation}
  \label{eq:hmat_elec}
  H_\text{mat}^\text{eff}
  = \sqrt{2} G_F N_e(r)
  \begin{pmatrix}
    -\Eps_D^e + c_{13}^2/2 & \Eps_N^e \\
    \Eps_N^{e*} & \Eps_D^e - c_{13}^2/2
  \end{pmatrix} .
\end{equation}
The probabilities are invariant under $H \to -H^*$, which is realized
for $\Dmq_{21} \to -\Dmq_{21}$ $\wedge$ $\left( \Eps_D^e - c_{13}^2/2
\right) \to -\left( \Eps_D^e - c_{13}^2/2 \right)$ $\wedge$ $\Eps_N^e
\to -\Eps_N^{e*}$. Combining this with the general symmetries
discussed above we can reabsorb the sign flip of both $\Dmq_{21}$ and
$\Eps_N^e$ into $\theta_{12}$, resulting in the transformation
$\theta_{12} \to \pi/2 - \theta_{12}$ $\wedge$ $\Eps_D^e \to c_{13}^2
- \Eps_D^e$ $\wedge$ $\Eps_N^e \to \Eps_N^{e*}$. This invariance
implies that for each point in the so-called ``light-side'' of the
parameter space (\textit{i.e.}, the region with $\theta_{12} <
45^\circ$) there is a point in the ``dark-side'' (the region with
$\theta_{12} > 45^\circ$) which cannot be distinguished experimentally
by oscillations alone.  In the case of NSI with $f=u$ or $f=d$ such a
symmetry is no longer exact, however as we will see in
Sec.~\ref{sec:solar} it is still realized with considerable accuracy.

As mentioned in Sec.~\ref{sec:form-atmos} the transition probabilities
in the atmospheric+LBL sector are invariant under a simultaneous sign
flip of $\Dmq_{31}$, $\Eps$ and $\alpha_i$. If this transformation is
extended to the solar+KamLAND sector through Eqs.~\eqref{eq:eps_atm},
\eqref{eq:eps_true} and then~\eqref{eq:eps_D}, \eqref{eq:eps_N} (with
$\dltCP = 0$ as in the atmospheric approximation) it leads to
$\Eps_D^f \to c_{13}^2 / Y_f - \Eps_D^f$ $\wedge$ $\Eps_N^f \to
-\Eps_N^{f*}$. For $f=e$ this transformation becomes an exact symmetry
if combined with a sign flip of $\Dmq_{21}$, as we have just seen.
However, for $f=u$ or $f=d$ such symmetry is only approximate, so that
the inclusion of solar data can (at least in principle) lift the sign
degeneracy between $\Dmq_{31}$ and $\Eps$.

\section{Analysis of solar and KamLAND data}
\label{sec:solar}

Let us start by presenting the results of the updated analysis of
solar and KamLAND experiments in the context of oscillations with the
generalized matter potential in Eq.~\eqref{eq:hmatsol}.  For KamLAND
we include the observed energy spectrum in the DS-1 and DS-2 data
sets~\cite{Gando:2010aa} with a total exposure of $3.49\times 10^{32}$
target-proton-year (2135 days).  In the analysis of solar neutrino
experiments we include the total rates from the radiochemical
experiments Chlorine~\cite{Cleveland:1998nv},
Gallex/GNO~\cite{Kaether:2010ag} and
SAGE~\cite{Abdurashitov:2009tn}. For real-time experiments we include
the 44 data points of Super-Kamiokande phase I (SK1) energy-zenith
spectrum~\cite{Hosaka:2005um}, the 33 data points of
SK2~\cite{Cravens:2008aa} and 42 data points of SK3~\cite{Abe:2010hy}
energy and day/night spectra, and the 24 data points of the 1097-day
energy spectrum and day-night asymmetry of SK4~\cite{Smy2012}.  We
also include the main set of the 740.7 days of Borexino
data~\cite{Bellini:2011rx} as well as their high-energy spectrum from
246 live days~\cite{Bellini:2008mr}.

The results of the three phases of SNO are included in two different
forms. First, we perform our own combined analysis of the 34 data
points of the day-night spectrum data of SNO-I~\cite{Aharmim:2007nv},
the 38 data points of the day-night spectrum of
SNO-II~\cite{Aharmim:2005gt} and the three total rates of
SNO-III~\cite{Aharmim:2008kc}. We label this analysis as
\textsc{SNO-data}. Second, we use the results of their the low energy
threshold analysis of the combined SNO phases
I--III~\cite{Aharmim:2011vm} which is given in the form of an
effective \emph{MSW-like} polynomial parametrization for the day and
night survival probabilities --~under the assumption of unitarity of
the oscillation probabilities~-- in terms of 7 parameters for which
the collaboration give the best fit values and covariant matrix. We
label this analysis as \textsc{SNO-poly}. Strictly the results of this
effective parametrization cannot be used for study of exotic scenarios
in which either unitarity in the active neutrino sector does not hold
(like for scenarios with sterile neutrinos) or the energy dependence
of the oscillation probability cannot be well represented by a simple
quadratic function.  Thus in order to verify the robustness of our
conclusions on the matter potential we present our results for both
variants of the SNO analysis. In both cases we have used the solar
fluxes from the Standard Solar Model GS98~\cite{Bahcall:2004pz,
  PenaGaray:2008qe}.

\begin{figure}\centering
  \includegraphics[width=\textwidth]{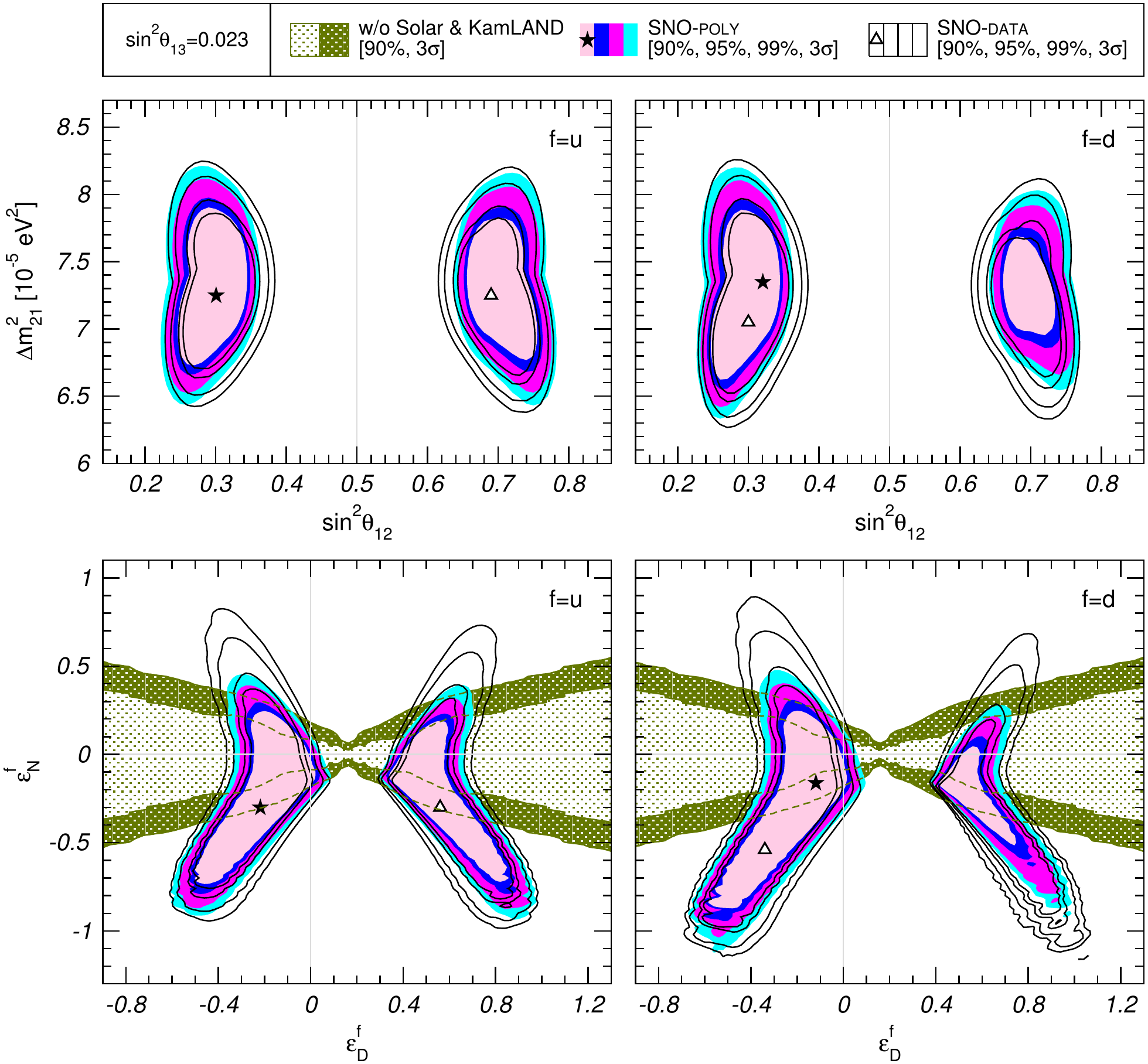}
  \caption{Two-dimensional projections of the 90\%, 95\%, 99\% and
    $3\sigma$ CL (2~dof) allowed regions from the analysis of solar
    and KamLAND data in the presence of non-standard matter
    potential. The results are shown for a fix value of
    $\sin^2\theta_{13}=0.023$ and after marginalizing over the two
    undisplayed parameters. The left (right) panels corresponds to
    $f=u$ ($f=d$).  The colored filled (black-contour void) regions in
    each panel correspond to the \textsc{SNO-poly} (\textsc{SNO-data})
    variants of the solar analysis, see text for details.  The best
    fit point is marked with a star (triangle).  For comparison we
    show also in the lower panels the two green dotted areas
    correspond to the 90\% and $3\sigma$ CL allowed regions from the
    analysis of the atmospheric and LBL data.}
  \label{fig:region-sun}
\end{figure}

\begin{figure}\centering
  \includegraphics[width=\textwidth]{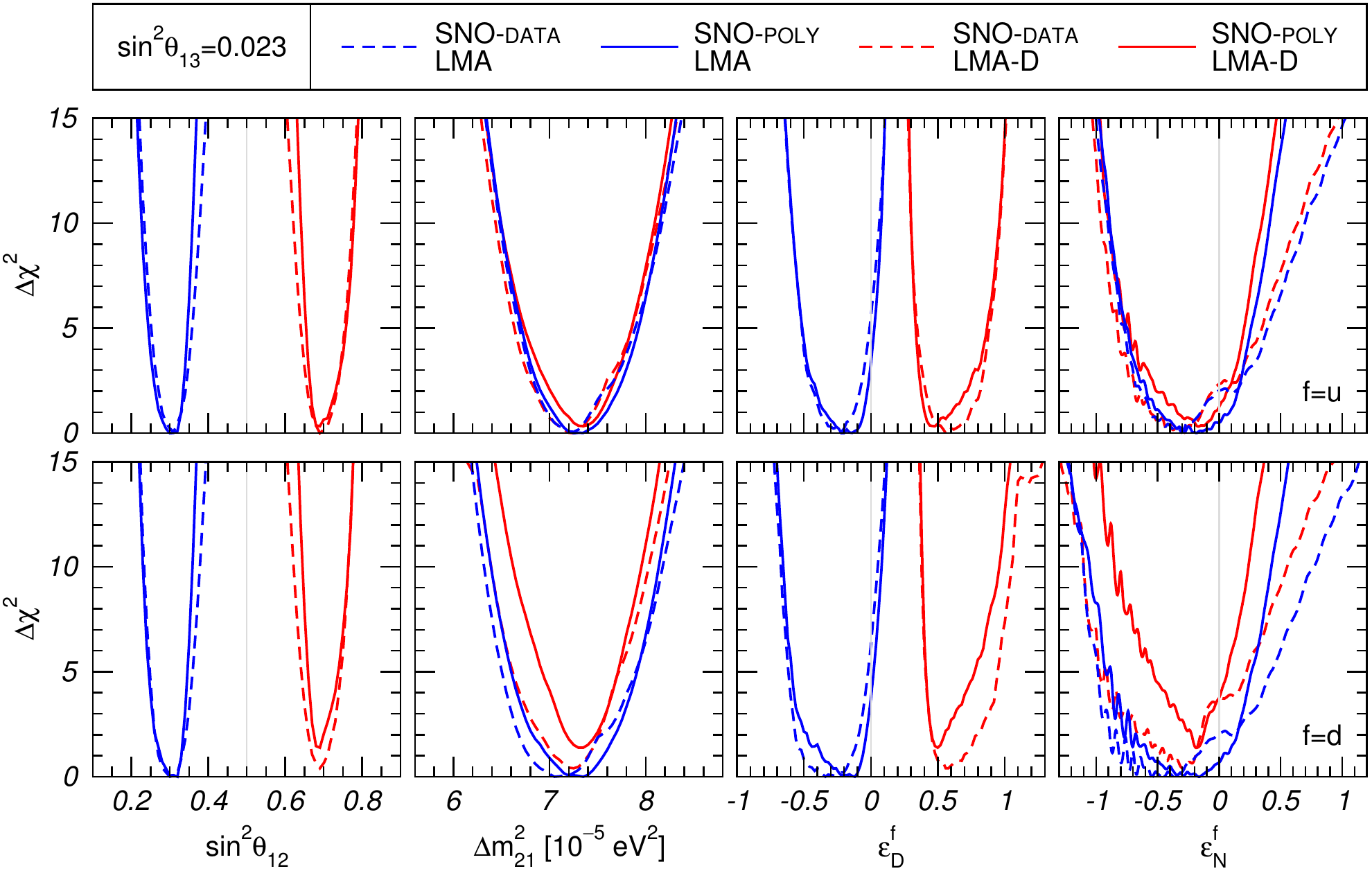}
  \caption{Dependence of the $\Delta\chi^2$ function for the analysis
    of the solar and KamLAND data on the relevant oscillation and
    matter potential parameters for $f=u$ (upper panels) and $f=d$
    (lower panels), for both LMA and LMA-D regions and the two
    variants of the SNO analysis, as labeled in the figure.}
  \label{fig:chisq-sun}
\end{figure}

We present the results of the analysis of solar and KamLAND data in
Figs.~\ref{fig:region-sun} and~\ref{fig:chisq-sun}.  The presence of
NSI with electrons, $f=e$, would affect not only neutrino propagation
in matter as described in Eq.~\eqref{eq:hmatsol}, but also the
neutrino-electron cross-section in experiments such as SK and
Borexino.  Since here we are only interested in studying the bounds to
propagation effects we will consider only the cases $f=u$ and $f=d$.
Also for simplicity the results are shown for real
$\Eps_N^f$. Strictly speaking, as discussed in
Sec.~\ref{sec:form-solar}, the sign of $\Eps_N^f$ is not physically
observable in oscillation experiments, as it can be reabsorbed into a
redefinition of the sign of $\theta_{12}$. However, for definiteness
we have chosen to present our results in the convention $\theta_{12}
\geq 0$, and therefore we consider both positive and negative values
of $\Eps_N^f$.

Fig.~\ref{fig:region-sun} shows the two-dimensional projections on the
oscillation parameters ($\Dmq_{21}$, $\sin^2\theta_{12}$) and the
matter potential parameters ($\Eps_N^f$, $\Eps_D^f$) with $f=u,d$
after marginalizing on the undisplayed parameters, for a fix value of
$\sin^2\theta_{13}=0.023$ which is the best fit for the global
analysis for $3\nu$ oscillations~\cite{GonzalezGarcia:2012sz,
  nufit-1.1}.
The first thing to notice is that for both \textsc{SNO-data} and
\textsc{SNO-poly} variants there are two disconnected regions in the
parameter space. The leftmost region in each panel, whose projection
on the oscillation parameters lies in the first octant of
$\theta_{12}$ ($0 \leq \theta_{12} \leq 45^\circ$) and whose
projection on the matter potential parameters contains SM case
(\textit{i.e.}, the point $\Eps_N^f = \Eps_D^f = 0$), corresponds to
the variation of the ``standard'' LMA solution in the presence of NSI,
so we will refer to it simply as LMA.  The rightmost region in each
panel, whose projection on the oscillation parameters lies the second
octant of $\theta_{12}$ ($45^\circ \leq \theta_{12} \leq 90^\circ$)
and whose projection on the matter potential does not contain the SM
point $\Eps_N^f = \Eps_D^f = 0$, corresponds to the ``dark-side''
solution found in Ref.~\cite{Miranda:2004nb} where it was labeled as
LMA-D. The existence of this new solution, almost degenerate with the
usual one, is consequence of the quasi-symmetry of the matter
potential discussed below Eq.~\eqref{eq:hmat_elec}.  We find that at
present the best fit point is in most of the cases in the LMA region,
but LMA-D lies only at a $\Delta\chi^2 = -0.06$ ($f=u$) and $0.4$
($f=d$) in the \textsc{SNO-data} variant, increasing to $\Delta\chi^2
= 0.3$ ($f=u$) and $1.4$ ($f=d$) in the \textsc{SNO-poly} variant.  As
seen in the lower panels the LMA-D solution requires a non-standard
matter potential with quite sizable values of $\Eps_D^f$.  An obvious
question is whether such large values are in contradiction with other
neutrino oscillation data, in particular with atmospheric neutrinos.
We will return quantitatively to this point in the next section but
for illustration we show also in the lower panels as dotted green
regions the corresponding 90\% and $3\sigma$ CL (2~dof) from the
analysis of atmospheric and LBL experiments. We see from the figure
that still a sizable fraction of the required non-standard matter
potential parameters for the LMA-D solution is compatible with all the
oscillation data.

In what respects the dependence on $f$ the figure shows that
although there are small quantitative differences, qualitatively the
results are rather similar for non-standard potential for $u$ or $d$
quarks. Also both variants of the SNO analysis yield similar results.

\begin{figure}\centering
  \includegraphics[width=\textwidth]{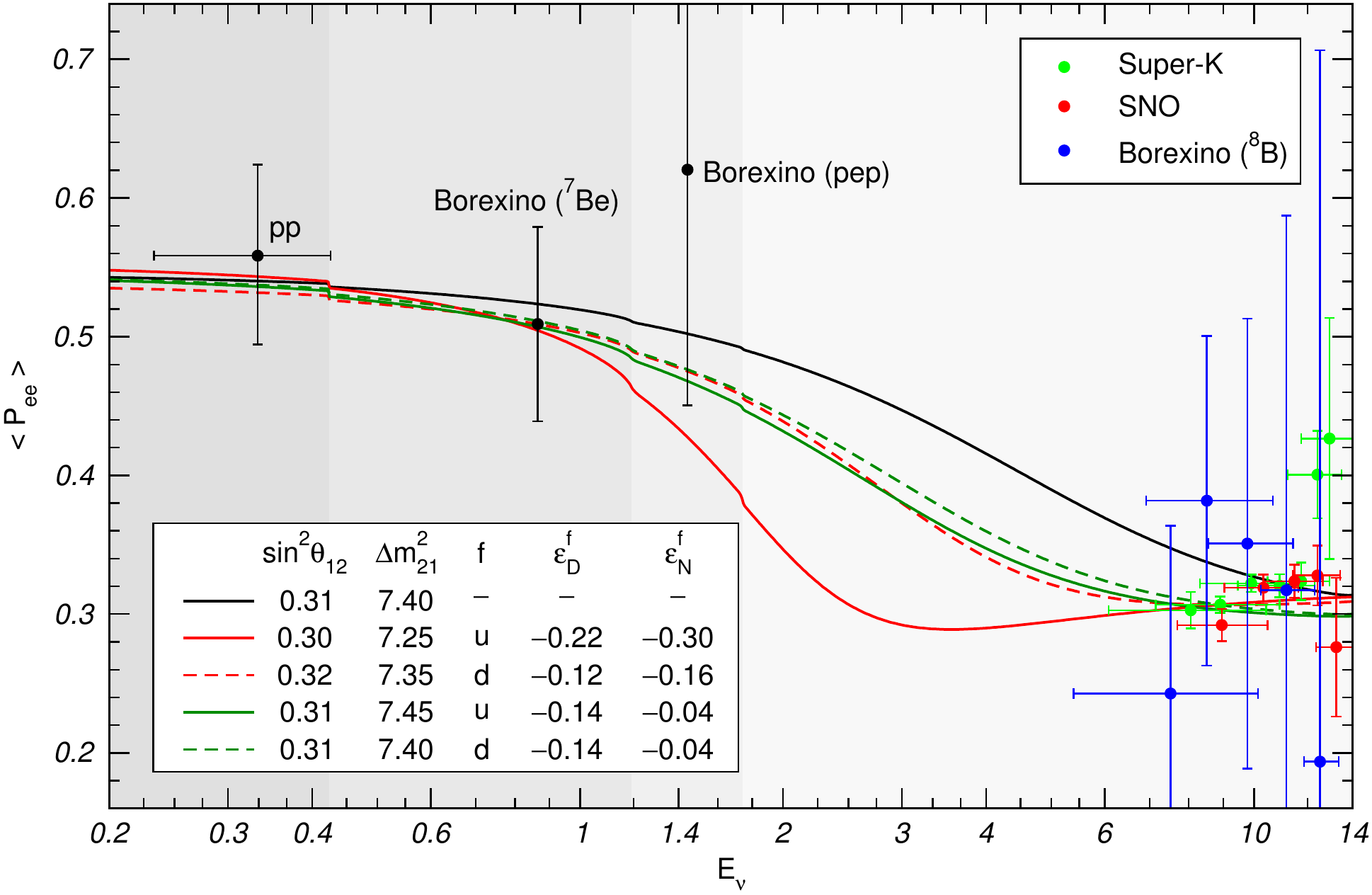}
  \caption{Survival probabilities in the Sun for different sets of
    oscillation and matter potential parameters as labeled in the
    figure.  In all cases we set $\sin^2\theta_{13} = 0.023$; the
    quoted value of $\Dmq_{21}$ is given in units of
    $10^{-5}~\eVq$. For illustration we also show the extracted
    average survival probabilities from different experiments. See
    text for details.}
  \label{fig:energy-sun}
\end{figure}

Fig.~\ref{fig:chisq-sun} contains the dependence of $\Delta\chi^2$ on
each of the four parameters $\Dmq_{21}$, $\theta_{12}$, $\Eps_D^f$,
$\Eps_N^f$, again for $\sin^2\theta_{13} = 0.023$ after marginalizing
over the other three. In each panel the four curves correspond to the
LMA and LMA-D solutions for both variants of the SNO analysis. The
main feature to notice is that in all cases the fit prefers some
non-standard value of the matter potential parameters, while for any
$f$ the SM potential lies at a $\Delta\chi^2 = 5.3$ and $\Delta\chi^2
= 7.9$ for \textsc{SNO-poly} and \textsc{SNO-data}, respectively. This
arises from the well-known fact that neither the SNO nor SK4 low
energy threshold analysis nor the \Nuc{8}{B} measurement in Borexino
seem to show evidence of the low energy turn-up of the spectrum
predicted in the standard LMA MSW solution.  This behavior can be
better described in the presence of a non-standard matter potential.
This is illustrated in Fig.~\ref{fig:energy-sun} where we show the
survival probability of solar neutrinos as a function of the neutrino
energy, for the best fit of oscillations only (black line) as well as
the best fits for $f=u$ and $f=d$ in the presence of NSI from the
analysis of solar+KamLAND data (red lines) and from the global
analysis discussed in the next section (green lines). In order to take
into account the dependence on the neutrino production point, which is
of particular relevance in the presence of non-standard matter
potential, we define the average survival probability $\langle P_{ee}
\rangle$ as
\begin{equation}
  \langle P_{ee} (E_\nu) \rangle
  = \dfrac{\sum_i \Phi_i(E_\nu) \int \rho_i(r) \, P_{ee}(E_\nu,r) \, dr}
  {\sum_i \Phi_i(E_\nu)}
\end{equation} 
where $i = \text{pp}$, pep, \Nuc{7}{Be}, \Nuc{13}{N}, \Nuc{15}{O},
\Nuc{17}{F}, \Nuc{8}{B} and hep labels the neutrino production
reaction and $\rho_i(r)$ is the distribution of production points for
the reaction $i$ normalized to 1.

\section{Results of global analysis}
\label{sec:global}

We now present the results of the global analysis including also
atmospheric, LBL and all other reactor data.  The data samples
included here are the same as in the NuFIT~1.1 analysis described in
Ref.~\cite{nufit-1.1}. For atmospheric data we use the
Super-Kamiokande results from phases 1--4~\cite{skatm1-4}, adding the
1097 days of phase 4 to their published data from phases
1--3~\cite{Wendell:2010md}.  For what concerns long-baseline
accelerator experiments, we combine the energy distribution obtained
by MINOS in both $\nu_\mu$ ($\bar\nu_\nu$)
disappearance~\cite{Adamson:2013whj} and $\nu_e$ ($\bar\nu_e$)
appearance with $10.7~(3.36) \times 10^{20}$ protons on
target~\cite{Adamson:2013ue}, and T2K $\nu_e$ appearance and $\nu_\mu$
disappearance data for phases 1--3 corresponding to $3.01\times
10^{20}$ pot~\cite{t2k:moriond13}.
For oscillation signals at reactor experiments, besides KamLAND, we
include data from the finalized experiments
CHOOZ~\cite{Apollonio:1999ae} (energy spectrum data) and Palo
Verde~\cite{Piepke:2002ju} (total rate) together with the recent
spectrum from Double Chooz with 227.9 days live
time~\cite{Abe:2012tg}, and the total even rates in the near and far
detectors in Daya Bay~\cite{An:2012bu} and Reno with 402 days of
data-taking~\cite{reno:venice13}.  For the reactor fluxes we follow
here the approach of Ref.~\cite{Schwetz:2011qt}, \textit{i.e.}, we
introduce an overall flux normalization which is then fitted to the
data together with the oscillation and matter potential parameters.
To better constrain such reactor flux normalization we also include in
the analysis the results of the reactor experiments
Bugey4~\cite{Declais:1994ma}, ROVNO4~\cite{Kuvshinnikov:1990ry},
Bugey3~\cite{Declais:1994su}, Krasnoyarsk~\cite{Vidyakin:1987ue,
  Vidyakin:1994ut}, ILL~\cite{Kwon:1981ua},
G\"osgen~\cite{Zacek:1986cu}, SRP~\cite{Greenwood:1996pb}, and
ROVNO88~\cite{Afonin:1988gx}, which due to their short baselines ($L
\lesssim 100$~m) are insensitive to the neutrino oscillation effects
discussed here.

\begin{figure}\centering
  \includegraphics[width=\textwidth]{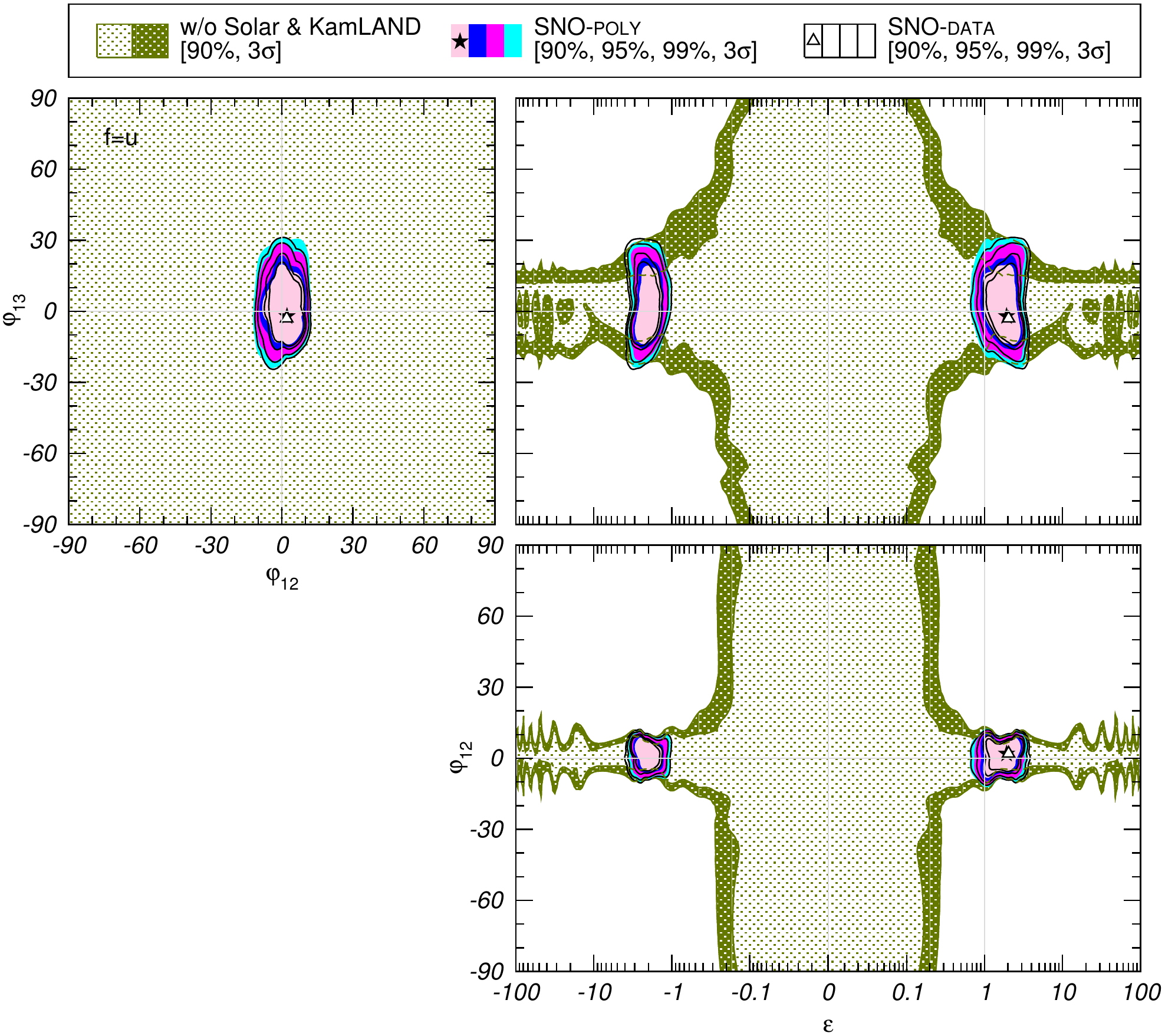}
  \caption{Two-dimensional projections on the matter potential
    parameters ($\Eps$, $\varphi_{12}$, $\varphi_{13}$) of the 90\%,
    95\%, 99\% and $3\sigma$ CL (2~dof) allowed regions from the
    global analysis of solar, atmospheric, reactor and LBL data after
    marginalization with respect to the undisplayed parameters.  The
    colored filled (black-contour void) regions in each panel
    correspond to $f=u$ and the \textsc{SNO-poly} (\textsc{SNO-data})
    variants of the solar analysis.  The best fit point is marked with
    a star (triangle).  For comparison we also show as green dotted
    areas the 90\% and $3\sigma$ CL regions from the analysis of
    atmospheric, LBL and reactor neutrinos (without solar nor
    KamLAND).}
  \label{fig:region-eps}
\end{figure}

\begin{figure}\centering
  \includegraphics[width=\textwidth]{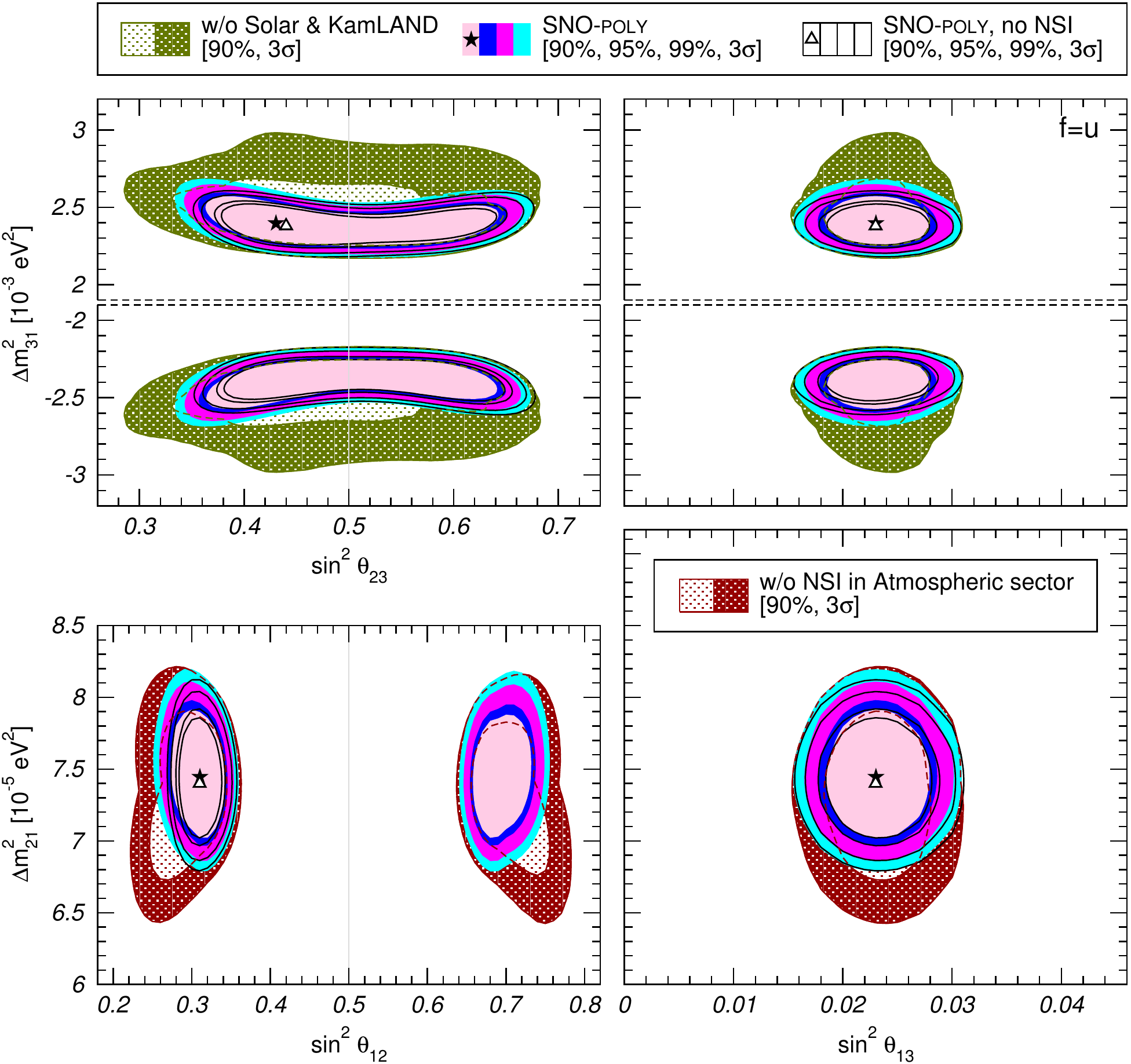}
  \caption{Two-dimensional projections of the 90\%, 95\%, 99\% and
    $3\sigma$ CL (2~dof) allowed regions of the oscillation parameters
    for $f=u$ and the \textsc{SNO-poly} variant of the solar analysis,
    after marginalizing over the matter potential parameters and the
    undisplayed oscillation parameters. The full regions and the star
    correspond to the global analysis including NSI, while the
    black-contour void regions and the triangle correspond to the
    analysis with the usual SM potential.  The green and red dotted
    areas show the 90\% and $3\sigma$ CL allowed regions from partial
    analyses where the effects of the non-standard matter potential
    have been neglected either in the solar+KamLAND (green) or in the
    atmospheric+LBL (red) sectors.}
  \label{fig:region-osc}
\end{figure}

\begin{figure}\centering
  \includegraphics[width=\textwidth]{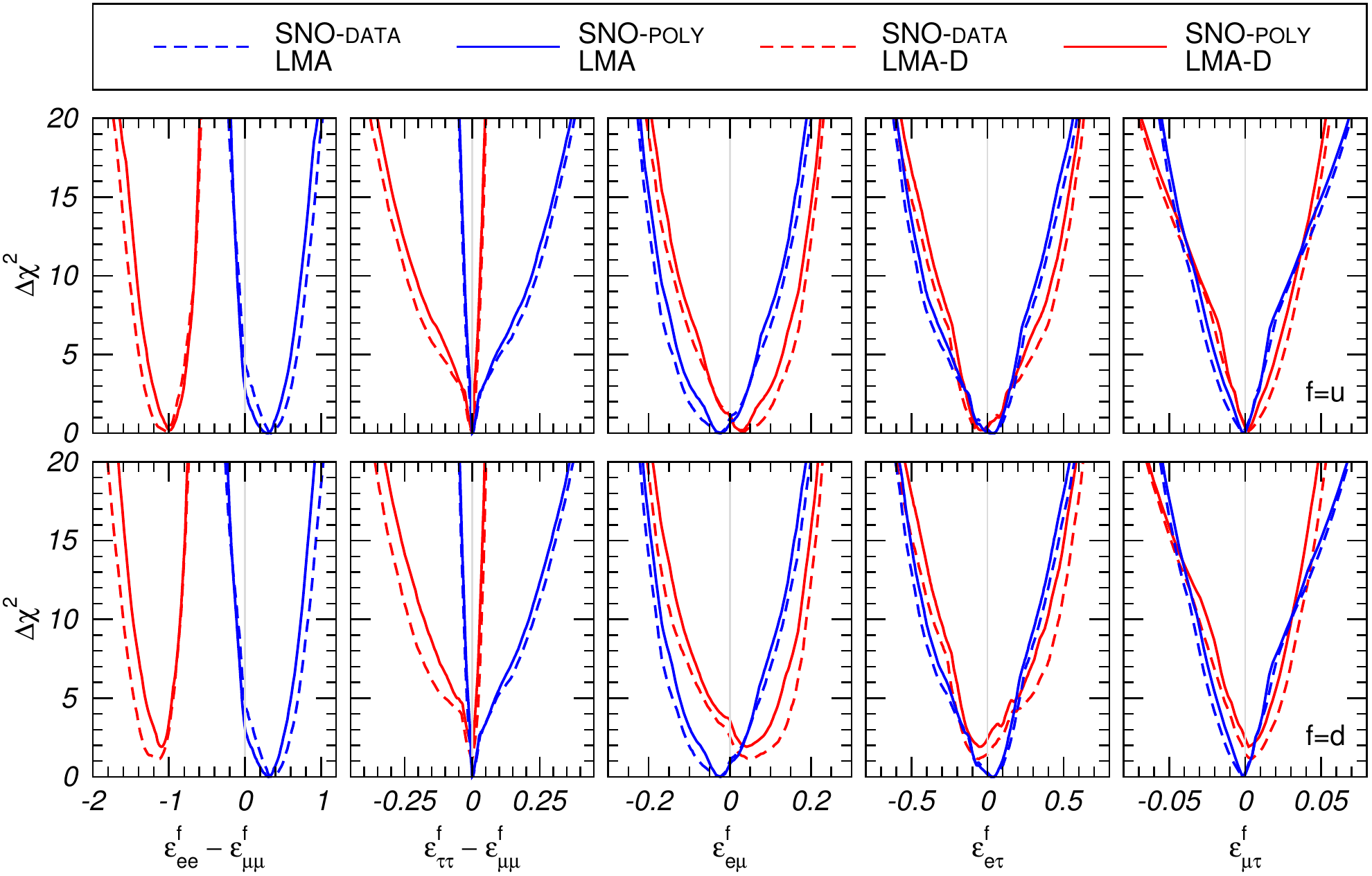}
  \caption{Dependence of the $\Delta\chi^2$ function for the global
    analysis of solar, atmospheric, reactor and LBL data on the NSI
    parameters $\Eps_{\alpha\beta}^f$ for $f=u$ (upper panels) and
    $f=d$ (lower panels), for both LMA and LMA-D regions and the two
    variants of the SNO analysis, as labeled in the figure.}
  \label{fig:chisq-eps}
\end{figure}

We present the results of the global analysis in
Figs.~\ref{fig:region-eps}, \ref{fig:region-osc}, and
\ref{fig:chisq-eps}.
In Fig.~\ref{fig:region-eps} we display the two-dimensional
projections of the allowed regions in the matter potential parameters
$\Eps$, $\varphi_{12}$ and $\varphi_{13}$ (in the parametrization of
Eq.~\eqref{eq:hmatgen} with the additional constraint of equal matter
eigenvalues $\Eps'=0$) after marginalizing over the oscillation
parameters $\Dmq_{21}$, $\Dmq_{31}$, $\theta_{12}$, $\theta_{23}$, and
$\theta_{13}$.  Since the $\alpha_i$ phases have little impact on our
results, we set for simplicity $\alpha_1 = \alpha_2 = 0$. Also, for
the sake of concreteness we focus here on $f=u$.
The filled colored (black-contour void) regions correspond to the
global analysis with the \textsc{SNO-poly} (\textsc{SNO-data}) variant
of the solar data.  For comparison we show also the dotted green
regions which correspond to the analysis of atmospheric, LBL and
reactor neutrinos (without solar nor KamLAND) and therefore update our
previous results of Ref.~\cite{GonzalezGarcia:2011my}.  As discussed
in Refs.~\cite{Friedland:2004ah, Friedland:2005vy,
  GonzalezGarcia:2011my} no bound on the magnitude of the matter
effects, $\Eps$, can be derived from the analysis of atmospheric and
LBL experiments in this general scenario. Specific bounds on $\Eps$
can be derived if a certain flavor structure of the matter potential
is assumed \emph{a priori} (for example, if we assume that no matter
effects are present in the $e\mu$ and $e\tau$ projections, which
corresponds to $\varphi_{12} = \pi/2$), implying that $\varphi_{12}$
and/or $\varphi_{13}$ are larger than some given value. Conversely
when marginalizing over $\Eps$ the full flavor projection
($\varphi_{12}$, $\varphi_{13}$) plane is allowed.  However, as seen
from the figure, once the results of solar and KamLAND experiments
(\textit{i.e.}, the samples involving $\nu_e$ or $\bar\nu_e$ and long
enough distances to see both oscillations and NSI effects) are
included in the analysis, a bound on the magnitude of the matter
effects $\Eps$ is obtained. Furthermore the flavor structure of the
potential is dramatically constrained as seen in upper-left panel.

Fig.~\ref{fig:region-osc} shows the two-dimensional projections of the
allowed regions from our global analysis in different combinations of
the oscillation parameters, again for $f=u$. The regions are obtained
after marginalizing over the undisplayed oscillation and matter
potential parameters. For comparison we also show as black-contour
void regions the corresponding results with the usual SM matter
potential.\footnote{Notice that in this analysis we are neglecting
  $\Dmq_{21}$ effects in the atmospheric and LBL oscillations, hence
  the standard oscillation results have no sensitivity to CP violation
  and only very marginal sensitivity to the mass ordering and the
  $\theta_{23}$ octant. For fully updated results and a complete
  treatment of neutrino oscillations in the standard case we address
  the reader to Refs.~\cite{GonzalezGarcia:2012sz, nufit-1.1}.}
The figure clearly shows the robustness of the determination of the
oscillation parameters even in the presence of a generalized matter
potential, with the exception of the octant of $\theta_{12}$.  In this
respect, we find that the LMA-D solution is still allowed in the
global analysis at $\Delta \chi^2 = 0.1$ ($0.2$) for $f=u$ and the
\textsc{SNO-data} (\textsc{SNO-poly}) variants, and at $\Delta\chi^2 =
1.1$ ($1.9$) for $f=d$ and the \textsc{SNO-data} (\textsc{SNO-poly})
analysis.  In the figure we also show as green or red dotted regions
the results of the analysis when the effects of the non-standard
matter parameters are neglected in either Solar+KamLAND (green, upper
panels) or in atmospheric+LBL (red, lower panels).  The comparison of
the global analysis with these partial analyses illustrates the
complementarity of the solar+KamLAND and the atmospheric+LBL data in
the robustness of the global fit. We also notice how in the upper
panels the green regions are perfectly symmetric under a sign flip of
$\Delta m^2_{31}$, as explained at the end of
Sec.~\ref{sec:form-atmos}.  However, for NSI with quarks ($f=u,d$)
this degeneracy is lifted once the solar data are also included in the
analysis, as discussed in Sec.~\ref{sec:form-solar}.  Thus the colored
regions are not exactly identical for both orderings, although with
present data the asymmetry is still minimal.

\begin{figure}\centering
  \includegraphics[width=\textwidth]{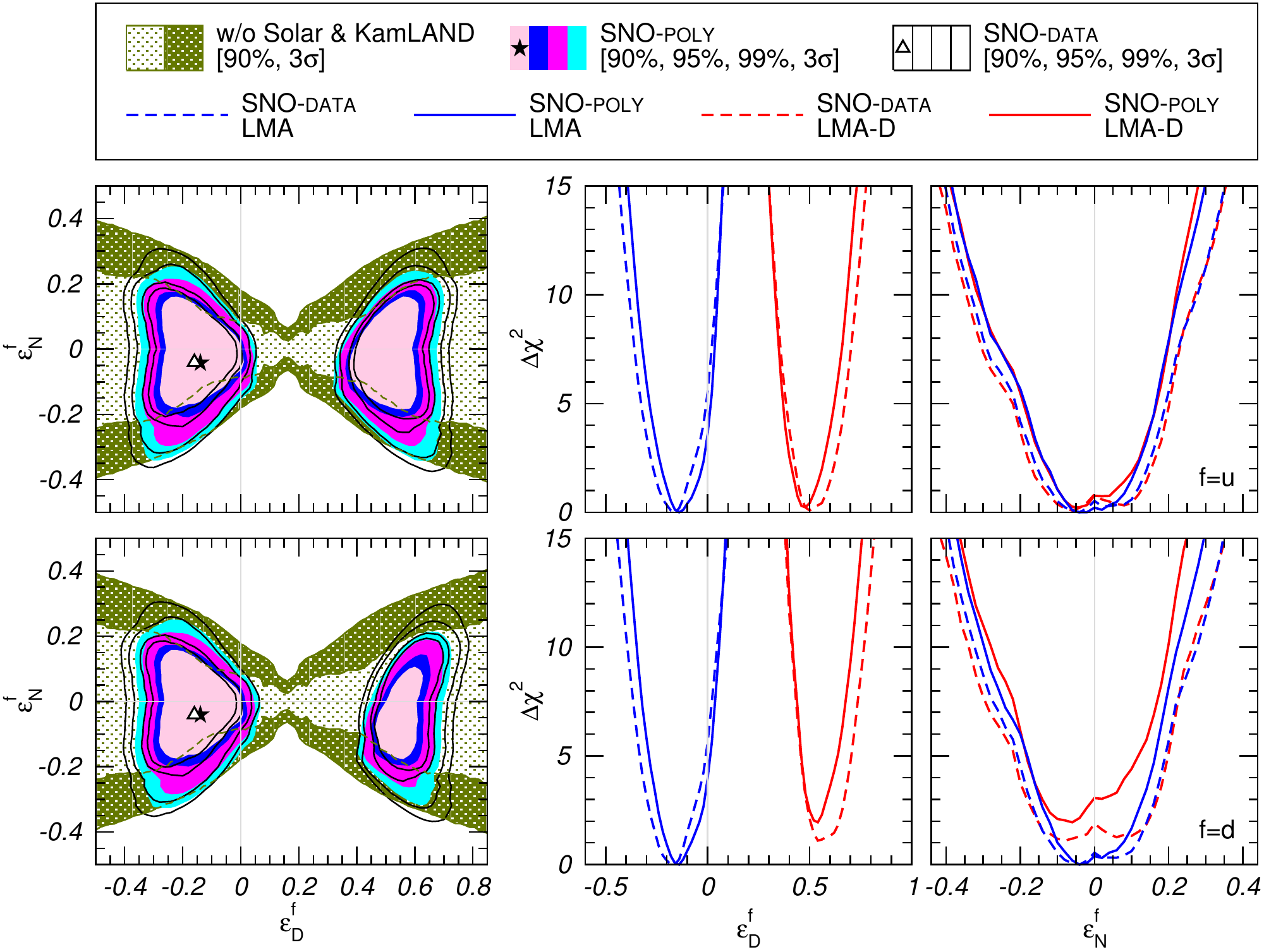}
  \caption{Constraints on the effective matter potential parameters
    $\Eps_D^f$ and $\Eps_N^f$ relevant in solar neutrino propagation
    for $f=u$ (upper panels) and $f=d$ (lower panels).  In the left
    panels we show as colored filled (black-contour void) areas the
    two-dimensional projections of the 90\%, 95\%, 99\% and $3\sigma$
    CL (2~dof) allowed regions from the global analysis, for the
    \textsc{SNO-poly} (\textsc{SNO-data}) variants of the solar
    analysis.  The best fit point is marked with a star (triangle).
    The green dotted areas correspond to the 90\% and $3\sigma$ CL
    allowed regions from the analysis of atmospheric, LBL and reactor
    data (without solar and KamLAND).  The central and right panels
    show the dependence of $\Delta\chi^2$ from the global analysis on
    $\Eps_D^f$ and $\Eps_N^f$, as labeled in the figure.}
  \label{fig:mixed-eps}
\end{figure}

In Fig.~\ref{fig:chisq-eps} we plot the dependence of the
$\Delta\chi^2$ function for the global analysis on the NSI parameters
$\Eps_{\alpha\beta}^f$, after marginalizing over the undisplayed
oscillation and matter potential parameters. Similarly, in
Fig.~\ref{fig:mixed-eps} we show the present determination on the
effective matter potential parameters $\Eps_D^f$ and $\Eps_N^f$
relevant in the propagation of solar and KamLAND neutrinos. In both
figures we display separately the results of the marginalization in
the LMA and the LMA-D regions of the parameter space, as well as both
the \textsc{SNO-data} and \textsc{SNO-poly} variants of the solar
analysis.
From these figures we derive the 90\% and $3\sigma$ allowed ranges for
the NSI parameters implied by the global analysis, which we summarize
in Table~\ref{tab:final}. The results in this table correspond to the
\textsc{SNO-poly} analysis and have been obtained for real matter
potential parameters. As discussed in Sec.~\ref{sec:formalism}, in
such a case only the \emph{relative} sign of the various
$\Eps_{\alpha\neq\beta}^f$ and the vacuum mixing angles can be
determined by oscillations. Thus strictly speaking once the results
are marginalized with respect to all other parameters in the most
general parameter space, the oscillation analysis can only provide
bounds on $|\Eps_{\alpha\neq\beta}^f|$. Still, for the sake of
completeness we have decided to retain in Table~\ref{tab:final} the
signs of the non-diagonal $\Eps_{\alpha\neq\beta}^f$, which is correct
as long as such signs are understood to be relative vacuum-matter
quantities and not intrinsic NSI features.

Neutrino scattering experiments such as CHARM~\cite{Dorenbosch:1986tb,
  Allaby:1987vr}, CDHSW~\cite{Blondel:1989ev} and
NuTeV~\cite{Zeller:2001hh} are sensitive to NSI with $u$ and $d$, and
can therefore yield information on
$\Eps_{\alpha\beta}^f$~\cite{Davidson:2003ha}. In
Ref.~\cite{Miranda:2004nb} it was found that the combination with
CHARM scattering results~\cite{Dorenbosch:1986tb, Allaby:1987vr} for
$f=d$ substantially lifts the statistical difference between LMA and
LMA-D. Although a rigorous combined analysis of the oscillation
results presented here with those from scattering experiments is
beyond the scope of this paper,\footnote{Notice that neutrino
  scattering results also depend on the axial NSI interactions and a
  rigorous global study of neutrino oscillation and scattering data
  will contain a larger number of parameters which makes it
  technically challenging.} in Table~\ref{tab:final} we present
separate ranges for marginalization over $0 \le \theta_{12} \le
45^\circ$ (denoted ``LMA'') and over the complete parameter space $0
\le \theta_{12} \le 90^\circ$ (denoted ``$\text{LMA} \oplus
\text{LMA-D}$''), so to give at least an idea of what could be gained
from scattering experiments. In most of the cases the $\text{LMA}
\oplus \text{LMA-D}$ marginalization yield just a slightly wider
interval than the marginalization within the LMA region.  However, for
$\Eps_{ee}^f - \Eps_{\mu\mu}^f$ and $\Eps_D^f$ the general allowed
range is composed by two separated intervals, one arising from the LMA
region and the other from the LMA-D region, so the full $\text{LMA}
\oplus \text{LMA-D}$ range has to be intended as the direct sum of the
bound provided in the LMA case and the extra interval quoted in the
$\text{LMA} \oplus \text{LMA-D}$ column.

\begin{table}\centering
  \definecolor{grey}{gray}{0.75}
  \newcommand{\grsep}{~\color{grey}\vrule\!\!}
  \begin{tabular}{|l|r|r@{\grsep}r|r@{\grsep}r|}
    \hline
    \multicolumn{2}{|c|}{}
    & \multicolumn{2}{c|}{90\% CL}
    & \multicolumn{2}{c|}{$3\sigma$}
    \\
    \hline
    Param. & best-fit
    & \multicolumn{1}{c@{\grsep}}{LMA}
    & \multicolumn{1}{c|}{$\text{LMA} \oplus \text{LMA-D}$}
    & \multicolumn{1}{c@{\grsep}}{LMA}
    & \multicolumn{1}{c|}{$\text{LMA} \oplus \text{LMA-D}$}
    \\
    \hline
    \hline
    $\Eps_{ee}^u - \Eps_{\mu\mu}^u$ & $+0.298$
    & $[+0.00, +0.51]$ & ${}\oplus [-1.19, -0.81]$
    & $[-0.09, +0.71]$ & ${}\oplus [-1.40, -0.68]$
    \\
    $\Eps_{\tau\tau}^u - \Eps_{\mu\mu}^u$ & $+0.001$
    & $[-0.01, +0.03]$ & $[-0.03, +0.03]$
    & $[-0.03, +0.20]$ & $[-0.19, +0.20]$
    \\
    $\Eps_{e\mu}^u$ & $-0.021$
    & $[-0.09, +0.04]$ & $[-0.09, +0.10]$
    & $[-0.16, +0.11]$ & $[-0.16, +0.17]$
    \\
    $\Eps_{e\tau}^u$ & $+0.021$
    & $[-0.14, +0.14]$ & $[-0.15, +0.14]$
    & $[-0.40, +0.30]$ & $[-0.40, +0.40]$
    \\
    $\Eps_{\mu\tau}^u$ & $-0.001$
    & $[-0.01, +0.01]$ & $[-0.01, +0.01]$
    & $[-0.03, +0.03]$ & $[-0.03, +0.03]$
    \\
    \hline
    $\Eps_D^u$ & $-0.140$
    & $[-0.24, -0.01]$ & ${}\oplus [+0.40, +0.58]$
    & $[-0.34, +0.04]$ & ${}\oplus [+0.34, +0.67]$
    \\
    $\Eps_N^u$ & $-0.030$
    & $[-0.14, +0.13]$ & $[-0.15, +0.13]$
    & $[-0.29, +0.21]$ & $[-0.29, +0.21]$
    \\
    \hline
    \hline
    $\Eps_{ee}^d - \Eps_{\mu\mu}^d$ & $+0.310$
    & $[+0.02, +0.51]$ & ${}\oplus [-1.17, -1.03]$
    & $[-0.10, +0.71]$ & ${}\oplus [-1.44, -0.87]$
    \\
    $\Eps_{\tau\tau}^d - \Eps_{\mu\mu}^d$ & $+0.001$
    & $[-0.01, +0.03]$ & $[-0.01, +0.03 ]$
    & $[-0.03, +0.19]$ & $[-0.16, +0.19]$
    \\
    $\Eps_{e\mu}^d$ & $-0.023$
    & $[-0.09, +0.04]$ & $[-0.09, +0.08 ]$
    & $[-0.16, +0.11]$ & $[-0.16, +0.17]$
    \\
    $\Eps_{e\tau}^d$ & $+0.023$
    & $[-0.13, +0.14]$ & $[-0.13, +0.14 ]$
    & $[-0.38, +0.29]$ & $[-0.38, +0.35]$
    \\
    $\Eps_{\mu\tau}^d$ & $-0.001$
    & $[-0.01, +0.01]$ & $[-0.01, +0.01 ]$
    & $[-0.03, +0.03]$ & $[-0.03, +0.03]$
    \\
    \hline
    $\Eps_D^d$ & $-0.145$
    & $[-0.25, -0.02]$ & ${}\oplus [+0.49, +0.57 ]$
    & $[-0.34, +0.05]$ & ${}\oplus [+0.42, +0.70]$
    \\
    $\Eps_N^d$ & $-0.036$
    & $[-0.14, +0.12]$ & $[-0.14, +0.12 ]$
    & $[-0.28, +0.21]$ & $[-0.28, +0.21]$
    \\
    \hline
  \end{tabular}
  \caption{90\% and $3\sigma$ allowed ranges for the matter potential
    parameters $\Eps_{\alpha\beta}^f$ for $f=u,d$ as obtained from the
    global analysis of oscillation data.  The results are obtained
    after marginalizing over oscillation and the other matter
    potential parameters either within the LMA only and within either
    LMA or LMA-D subspaces respectively.  The numbers quoted are the
    \textsc{SNO-poly} variant of the solar analysis. See text for
    details.}
  \label{tab:final}
\end{table}

\section{Summary}
\label{sec:summary}

In this article we have quantified our current knowledge of the size
and flavor structure of the matter background effects in the evolution
of solar, atmospheric, reactor and LBL neutrinos based solely on a
global analysis of oscillation data.  It complements the study in
Ref.~\cite{GonzalezGarcia:2011my} where the analysis of the matter
potential was perform only considering atmospheric and LBL neutrinos.

After briefly presenting the most general parametrization of the
matter potential and its connection with non-standard neutrino
interactions (NSI), we have focused on the analysis of solar and
KamLAND data. We have found (see Fig.~\ref{fig:chisq-sun}) that the
fit always prefers some non-standard value of the matter potential
parameters, while the SM potential lies at a $\Delta\chi^2 \sim
5$--$8$ depending on the details of the analysis. This is consequence
of the fact that none of the experiments sensitive to \Nuc{8}{B}
neutrinos has provided so far evidence of the low energy turn-up of
the spectrum predicted in the standard LMA MSW solution (see
Fig.~\ref{fig:energy-sun}). We have also found in that the present
analysis still allows for two disconnected regions in the parameter
space, the ``standard'' LMA region and the ``dark side'' LMA-D (see
Fig.~\ref{fig:region-sun}), and that the statistical difference
between both solutions never exceeds $\Delta\chi^2 = 1.4$.  Although
the LMA-D solution requires rather large values of the matter
parameters, we have shown (and latter quantified in
Sec.~\ref{sec:global}) that it is still fully compatible with the
bounds from atmospheric and LBL oscillation data.

We have then turned to a global analysis in which the data from solar
and KamLAND have been combined with those from atmospheric, LBL, and
other reactor experiments.  For what concerns the impact of the
non-standard matter potential on the determination of the oscillation
parameters, we found that the determination of $\Dmq_{21}$,
$|\Dmq_{31}|$, $\sin^2\theta_{23}$, and $\sin^2\theta_{13}$ is very
robust due to strong synergies between solar+KamLAND and
atmospheric+LBL data. In particular, once the results of solar and
KamLAND experiments are included in the analysis both the magnitude
and the flavor structure of NSI are strongly constrained, thus
preventing the weakening of the $|\Dmq_{31}|$ and $\theta_{23}$ bounds
which was observed in Refs.~\cite{Friedland:2004ah, Friedland:2005vy,
  GonzalezGarcia:2011my} from the analysis of atmospheric and LBL data
alone.  In turn, the inclusion of atmospheric+LBL data in the solar
analysis severely constrain the allowed range of non-diagonal NSI
described by the effective parameter $\Eps_N^f$, resulting in the
stabilization of the $\Dmq_{21}$ and $\sin^2(2\theta_{12})$
bounds. However, unlike for the case of oscillations with the usual SM
potential, in the presence of non-standard interactions a new solution
with $\sin^2\theta_{12} > 0.5$ (LMA-D region) becomes allowed.  With
all this, the $3\sigma$ ranges of the oscillation parameters read:
\begin{equation}
  \label{eq:oscrang}
  \begin{aligned}
    & \hspace{-10mm}\text{Standard Matter Potential}
    & \hspace{18mm}
    & \hspace{-10mm}\text{Generalized Matter Potential}
    \\[1mm]
    \sin^2\theta_{12} &\in [0.27, 0.35] \,,
    & \sin^2\theta_{12} &\in [0.26, 0.35] \oplus [0.65, 0.75] \,,
    \\
    \sin^2\theta_{23} &\in [0.36, 0.67] \,,
    & \sin^2\theta_{23} &\in [0.34, 0.67] \,,
    \\
    \sin^2\theta_{13} &\in [0.016, 0.030] \,,
    & \sin^2\theta_{13} &\in [0.016,0.030] \,,
    \\
    \Dmq_{21} &\in [6.87, 8.03] \times 10^{-5}~\eVq,
    & \Dmq_{21} &\in [6.86, 8.10] \times 10^{-5}~\eVq,
    \\
    |\Dmq_{31}| &\in [2.20, 2.58] \times 10^{-3}~\eVq,
    & |\Dmq_{31}| &\in [2.20, 2.65] \times 10^{-3}~\eVq.
  \end{aligned}
\end{equation}
The corresponding bounds on the individual NSI parameters from the
global analysis \emph{after marginalization from all other oscillation
  and matter parameters} are given in Fig.~\ref{fig:chisq-eps} and
Table~\ref{tab:final}.  Comparing the results in the Table with the
bounds derived in Refs.~\cite{Davidson:2003ha, Biggio:2009nt} from
non-oscillation data we find that, with the possible exception of
$\Eps_{e\mu}^{u,d}$, the global oscillation analysis presented here
yields the most restrictive bounds on the \emph{vector} NSI
parameters. This is even more impressive if one considers that the
one-dimensional bounds in Table~\ref{tab:final} arise as projections
of a global scan of the entire parameter space, and therefore
correlations among different parameters are properly take into
account. Conversely, the bounds from neutrino scattering experiments
are usually obtained on a one-by-one basis, \textit{i.e.}\ varying a
single parameter at a time while keeping all the others set to
zero. In spite of this, neutrino scattering experiments still provide
complementary information to oscillation experiments, for example for
$f=d$ they can substantially lifts the degeneracy between the LMA and
LMA-D solutions. Therefore, although a rigorous combined analysis of
neutrino oscillations and neutrino scattering experiments is
technically challenging and well beyond the scope of the present work,
it is certainly worth considering for the future.

\section*{Acknowledgments}

This work is supported by Spanish MINECO (grants FPA-2010-20807,
FPA-2012-31880, FPA-2012-34694, consolider-ingenio 2010 grant
CSD-2008-0037 and ``Centro de Excelencia Severo Ochoa'' program
SEV-2012-0249), by CUR Generalitat de Catalunya (grant 2009SGR502), by
Comunidad Autonoma de Madrid (HEPHACOS project S2009/ESP-1473), by
USA-NSF (grant PHY-09-6739) and by the European Union (EURONU project
FP7-212372 and FP7 Marie Curie-ITN actions PITN-GA-2009-237920
``UNILHC'' and PITN-GA-2011-289442 ``INVISIBLES'').

\bibliographystyle{JHEP}
\bibliography{references}

\end{document}